\documentclass[prd,twocolumn,amsmath,amssymb,superscriptaddress,nofootinbib]{revtex4}

\usepackage{graphicx}
\usepackage{dcolumn}
\usepackage{bm}

\normalbaselines \topmargin -0.25 truein \oddsidemargin 0.30
truein \evensidemargin 0.30 truein 

\newcommand{\D}{\hat{D}}

\begin{document}

\title{Spontaneous color polarization as a {\it modus originis} of the dynamic aether}

\author{Alexander B. Balakin}
\email{Alexander.Balakin@kpfu.ru} \affiliation{Department of General
Relativity and Gravitation, Institute of Physics, Kazan Federal University, Kremlevskaya
str. 18, Kazan 420008, Russia}
\author{Gleb B. Kiselev}
\email{gleb@karnaval.su} \affiliation{Department of General
Relativity and Gravitation, Institute of Physics, Kazan Federal University, Kremlevskaya
str. 18, Kazan 420008, Russia}

\date{\today}

\begin{abstract}
We suggest the phenomenological model of emergence of the dynamic aether as a result of decay of the SU(N) symmetric field configuration containing the multiplet of vector fields. The scenario of the transition to the dynamic aether, which is characterized by one unit timelike vector field associated with the aether velocity, is based on the idea of spontaneous color polarization analogous to the spontaneous electric polarization in ferroelectric materials. The mechanism of spontaneous color polarization is described in the framework of anisotropic cosmological model of the Bianchi-I type; it involves into consideration the idea of critical behavior of the eigenvalues of the tensor of color polarization in the course of the Universe accelerated expansion. The interim stage of transition from the color aether to the canonic dynamic aether takes the finite period of time, the duration of which is predetermined by the phenomenologically introduced critical value of the expansion scalar.
\end{abstract}

\maketitle

\section{Introduction}\label{Intro}

The term {\it dynamic aether} was introduced in the framework of the Einstein-aether theory  \cite{J1,J2,J3,J4,J5,J6,J7,J8,J9,J10}, which belongs to the vector-tensor branch of the science entitled Modified Theories of Gravity (see, e.g., \cite{O1,O2}). The theory of the dynamic aether operates with a timelike unit dynamic vector field $U^i$, which can be interpreted as the velocity four-vector of the cosmic substratum indicated as the aether. The interpretation of the vector field $U^i$ as the aether velocity four-vector revives inevitably the discussions concerning the
idea of a preferred frame of reference \cite{CW,N1,N2}, and the taboo associated with the Lorentz invariance violation \cite{LIV1,LIV2,LIV3}. One can interpret the appearance of the unit vector field $U^i$ using another idea: if there exists a timelike vector field ${\cal{U}}^i$ as a source of the gravity field, one can introduce the scalar field as the modulus of this vector field, ${\cal U}= \sqrt{g_{mn}{\cal{U}}^m{\cal{U}}^n}$, and the normalized unit four-vector $V^i \equiv \frac{{\cal{U}}^i}{{\cal{U}}}$. Thus, we deal with a version of the scalar-vector-tensor theory of gravity, based on the interacting trio $\left\{{\cal U}, V^j, g_{ik} \right\}$.

In the paper \cite{COLOR} we proposed an idea of the color aether, which is based on the introduction of the SU(N) symmetric multiplet of vector fields $U^i_{(a)}$. The subscript $(a)$ denotes the  group index in the adjoint representation of the SU(N) group and this Latin index runs over the values $\left\{(1), (2), ..(N^2-1)\right\}$. If all the vectors $U^i_{(a)}$ from this multiplet become parallel, i.e., when $U^i_{(a)}= q_{(a)} U^i$, and if $U^i$ is the unit timelike vector field, we happen to be faced with the version of the dynamic aether, supplemented by the SU(N)  symmetric multiplet of scalar fields $\{q_{(a)}\}$. Clearly, $q_{(a)}$ are the scalars from the point of view of spacetime transformations, but the corresponding object with upper index, $q^{(a)}= G^{(a)(b)}q_{(b)}$, can be considered as vectors in the $N^2{-}1$ dimensional group space with the metric $G^{(a)(b)}$ (color vectors, for short). When we deal with the SU(N) symmetric multiplet of vector fields, we have  to use the self-consistent formalism, which includes the Yang-Mills gauge fields with the potentials $A_k^{(a)}$, the gauge covariant derivative $D_m^{(a)}$, etc. This full-format SU(N) symmetric formalism is elaborated and presented in the work \cite{COLOR} as the extension of the U(1) symmetric formalism discussed in \cite{U1,U2,U3} in the framework of Einstein-Maxwell-aether theory.

In order to establish the phenomenological model of the transition from the color aether to the dynamic aether, we exploit the known analogy with the spontaneous polarization in the ferroelectric materials \cite{ferro,LL}; the term {\it spontaneous} means that the electric polarization is induced by variations of temperature or pressure, not by the external electric field. We keep in mind three principal details appeared in both models. The first detail is connected with the character of evolution. In the theory of ferroelectricity one can associate the evolution with the temperature decrease: high temperature relates to the symmetric phase without polarization, low temperature corresponds to the polarized dissymmetric phase. When we consider the evolution of the color system in the expanding Universe, the effective temperature is also dropping providing the transition from the symmetric phase to the dissymmetric one. The second detail is connected with the Curie temperature $T_C$: below this critical value, $T<T_C$, the spontaneous electric polarization appears, and thus, $T_C$ indicates the temperature of the ferroelectric phase transition. Similarly, we can introduce the critical value of the expansion scalar $\Theta_*$, which relates to the cosmological phase transition taking place at the cosmological time $t_*$. The third formal accordance is based on the analogy between electric domain structure in ferroelectric (chaotic in the symmetric phase, and aligned in the dissymmetric one) and the anisotropy of the group space associated with the color vector fields (multi-axial structure in the symmetric phase, and uni-axial in the dissymmetric one).
In other words, we assume that in the early Universe the phase transition took place, which transformed the SU(N) symmetric multiplet of vector fields to the bundle of vectors, which happen to be parallel in the group space. Below we describe the mathematical details of the phenomenological model of the spontaneous color polarization.

This paper is organized as follows. In Section II we recall the basic elements of the SU(N) symmetric field formalism and the corresponding master equations.
In Section III  we discuss the possible mechanism of transformation of the color aether, equipped by the SU(N) symmetric multiplet of the vector fields $\left\{ U^i_{(a)} \right\}$, to the canonic  dynamic aether, which possesses one unit vector field $U^i$. Section IV contains the discussion and conclusion.

\section{The formalism}

\subsection{Basic elements of the theory}

We follow the book \cite{Rubakov} and use the Hermitian traceless generators of the SU(N) group, ${\bf t}_{(a)}$, which satisfy
the commutation relations
\begin{equation}
\left[ {\bf t}_{(a)} , {\bf t}_{(b)} \right] = i  f^{(c)}_{\
(a)(b)} {\bf t}_{(c)} \,, \label{fabc}
\end{equation}
where $f^{(c)}_{\ (a)(b)}$ are the structure constants of the SU(N) gauge group.
The scalar product of the generators ${\bf t}_{(a)}$ and ${\bf t}_{(b)}$ introduces the metric in the group space, $G_{(a)(b)}$:
\begin{equation}
G_{(a)(b)} = \left( {\bf t}_{(a)} , {\bf t}_{(b)} \right) \equiv 2 {\rm Tr} \
{\bf t}_{(a)} {\bf t}_{(b)}    \,.
\label{scalarproduct}
\end{equation}
The structure constants satisfy the Jacobi identity
\begin{equation}
f^{(a)}_{\ (b)(c)} f^{(c)}_{\ (e)(h)}  {+} f^{(a)}_{\ (e)(c)} f^{(c)}_{\ (h)(b)} {+} f^{(a)}_{\ (h)(c)} f^{(c)}_{\ (b)(e)} {=} 0. \label{jfabc}
\end{equation}
The quantities
$$
f_{(c)(a)(b)} \equiv G_{(c)(d)} f^{(d)}_{\ (a)(b)} =
$$
\begin{equation}
= {-} 2 i
{\rm Tr}  \left[ {\bf t}_{(a)},{\bf t}_{(b)} \right] {\bf
t}_{(c)}   \label{fabc1}
\end{equation}
are antisymmetric with respect to transposition of any two indices.
The special (canonic) basis ${\bf t}_{(a)}$ provides the relations
\begin{equation}
\frac{1}{N} f^{(d)}_{\ (a)(c)} f^{(c)}_{\ (d)(b)} =
\delta_{(a)(b)} = G_{(a)(b)} \,.   \label{ff}
\end{equation}
The Yang-Mills field potential ${\bf A}_m$ and the Yang-Mills field strength ${\bf F}_{ik}$ are defined as follows:
\begin{equation}
{\bf A}_m = - i {\cal G} {\bf t}_{(a)} A^{(a)}_m \,, \quad {\bf F}_{mn} = - i {\cal G} {\bf t}_{(a)} F^{(a)}_{mn}  \,, \label{represent}
\end{equation}
where ${\cal G}$ is the coupling constant, and the multiplets of real fields  $A^{(a)}_i$ and $F^{(a)}_{ik}$ are connected as follows:
\begin{equation}
F^{(a)}_{mn} = \nabla_m
A^{(a)}_n - \nabla_n A^{(a)}_m + {\cal G} f^{(a)}_{\ (b)(c)}
A^{(b)}_m A^{(c)}_n \,. \label{46Fmn}
\end{equation}
Here and below $\nabla _m$ is a  covariant derivative.
The tensors $F^{ik}_{(a)}$ has its dual elements
\begin{equation}
{}^*\! F^{ik}_{(a)} = \frac{1}{2}\epsilon^{ikls} F_{ls (a)} \,,
\label{dual}
\end{equation}
with universal Levi-Civita tensor $\epsilon^{ikls}$.
The multiplet of vector fields $U^{(a)}_m$ appears as the decomposition of the Hermitian quantity
\begin{equation}
{\bf U}_m =  {\bf t}_{(a)} U^{(a)}_m \,.  \label{Urepresent}
\end{equation}
Our additional ansatz  is that the vector fields satisfy the condition
\begin{equation}
G_{(a)(b)} U^{(a)}_m U^{(b)}_n g^{mn} = 1 \,. \label{4represent}
\end{equation}
This is the direct generalization of the normalization condition $g^{mn}U_mU_n=1$ in the canonic Einstein-aether theory.
The extended (gauge covariant) derivative $\hat{D}_m$ is defined as follows:
$$
\D_m Q^{(a) \cdot \cdot \cdot}_{\cdot \cdot \cdot (d)} \equiv
\nabla_m Q^{(a) \cdot \cdot \cdot}_{\cdot \cdot \cdot (d)} + {\cal
G} f^{(a)}_{\cdot (b)(c)} A^{(b)}_m Q^{(c) \cdot \cdot
\cdot}_{\cdot \cdot \cdot (d)} -
$$
\begin{eqnarray}
- {\cal G} f^{(c)}_{\cdot (b)(d)}
A^{(b)}_m Q^{(a) \cdot \cdot \cdot}_{\cdot \cdot \cdot (c)} +...
\,, \label{DQ2}
\end{eqnarray}
where $Q^{(a) \cdot \cdot \cdot}_{\cdot \cdot \cdot (d)}$ is arbitrary tensor in the group space (see, e.g., \cite{Akhiezer}).
For the vector fields this formula gives
\begin{equation}
\D_m U^{(a)}_n \equiv \nabla_m
U^{(a)}_n + {\cal G} f^{(a)}_{\ (b)(c)} A^{(b)}_m U^{(c)}_n
\,. \label{DU}
\end{equation}
The metric $G_{(a)(b)}$ and the
structure constants $f^{(d)}_{\ (a)(c)}$ are the gauge covariant constant tensors in the group space:
\begin{equation}
\D_m G_{(a)(b)} = 0 \,, \qquad
\D_m
f^{(a)}_{\ (b)(c)} = 0 \,. \label{DfG}
\end{equation}
The dual tensor ${}^*\! F^{ik}_{(a)}$, due to the definitions (\ref{dual}) and (\ref{46Fmn}),
 satisfies the relations
\begin{equation}
\hat{D}_k {}^*\! F^{ik}_{(a)} = 0 \,. \label{Aeq2}
\end{equation}
Finally, we introduce the multiplet of scalar fields  $\Omega^{(a)}$, which forms the color vector
$$
\Omega^{(a)} \equiv \D_m U^{(a)m} =
$$
\begin{equation}
= \nabla_m
U^{(a)m} + {\cal G} f^{(a)}_{\ (b)(c)} A^{(b)}_m U^{(c)m} \,,
\label{A88}
\end{equation}
and the scalar $\Omega$ based on the relationship
\begin{equation}
\Omega^2 = \Omega^{(a)} \Omega_{(a)} \,. \label{A77}
\end{equation}
In the Abelian model of the aether with U(1) symmetry the scalar $\Omega$ coincides with the expansion scalar $\Theta \equiv \nabla_m U^m$.

\subsection{Action functional and master equations}

We use the action functional, which describes interaction of gauge, vector and gravitational fields
$$
S {=} \int d^4 x \sqrt{{-}g} \left\{ \frac{1}{2\kappa}\left[R {+} 2\Lambda {+} \lambda \left(U^m_{(a)}U^{(a)}_m {-}1 \right) {+} \right. \right.
$$
\begin{equation}\label{SUNact}
\left. \left. + {\cal K}^{ijmn}_{(a)(b)}\hat{D}_i U^{(a)}_{m}\hat{D}_j
U^{(b)}_n \right] {+} \frac{1}{4} F^{(a)}_{mn} F^{mn}_{(a)} \right\} \,.
\end{equation}
The full-format version of the constitutive tensor ${\cal K}^{ijmn}_{(a)(b)}$ is presented  in the paper \cite{COLOR}; that version contains 75 coupling constants.  Here, based on the direct analogy with the canonic Einstein-aether theory, we postulate the constitutive tensor to contain only four coupling constants:
$$
{\cal K}^{ijmn}_{(a)(b)} = C_{4} U^{i}_{(a)} U^{j}_{(b)} g^{mn} +
$$
\begin{equation}\label{constitutive1}
+ G_{(a)(b)} \left[C_1 g^{ij} g^{mn} {+} C_2 g^{im}g^{jn}
{+} C_3 g^{in}g^{jm} \right].
\end{equation}
Clearly, for the  U(1) - symmetric theory the group indices disappear, and this constitutive tensor coincides with the Jacobson's one \cite{J1}.

\subsubsection{Master equations for the gauge fields}

Master equations appear as a result of variation of the action functional (\ref{SUNact}) with respect to four quantities: the Lagrange multiplier $\lambda$, the vector fields $U^j_{(a)}$, the gauge potential four-vector $A_i^{(b)}$, and space-time metric $g^{pq}$. The variation with respect to $\lambda$ gives the normalization condition (\ref{4represent}).
Variation with respect to $A^{(a)}_i$ gives the extended Yang-Mills equations
\begin{equation}\label{Col1}
\hat{D}_k F^{ik}_{(a)} = \Gamma^i_{(a)}\,,
\end{equation}
where the color current $\Gamma^i_{(a)}$ is given by the formula
\begin{equation}\label{Col4}
 \Gamma^i_{(a)} =  \frac{{\cal G}}{\kappa}   f^{(d)}_{\ (c)(a)} U^{(c)}_k  {\cal K}^{imkn}_{(d)(b)} \hat{D}_m U^{(b)}_n  \,.
\end{equation}
This quantity appears, when we fulfil the variation of the second term in the right-hand side of (\ref{DU}) with respect to $A_i^{(d)}$. From the physical point of view, the color current $\Gamma^i_{(a)}$ is induced by the interaction between the color vector field and non-Abelian gauge field; clearly, it disappears in the U(1) case, since the structure constants are equal to zero in that model.

\subsubsection{Master equations for the vector fields $U^k_{(a)}$}

The variation with respect to $U^j_{(a)}$ yields the standard balance equation
\begin{equation}
\D_i {\cal J}^{ij}_{(a)}
 = \lambda \ U^j_{(a)}  +  {\cal I}^{j}_{(a)} \,,
\label{CU1}
\end{equation}
where the color objects  ${\cal J}^{ij}_{(a)}$ and ${\cal I}^j_{(a)}$ are introduced in analogy with the canonic aether theory \cite{J1}:
$$
{\cal J}^{ij}_{(a)} = {\cal K}^{imjn}_{(a)(b)} \hat{D}_m U^{(b)}_n  =
$$
$$
= C_1 \hat{D}^i U^{j}_{(a)} + C_2 g^{ij} \hat{D}_m U^{m}_{(a)} + C_3 \hat{D}^j U^{i}_{(a)}  +
$$
\begin{equation}
+ C_4 U^i_{(a)} U^m_{(b)} \hat{D}_m U^{j (b)}  \,,
\label{CU2}
\end{equation}
\begin{equation}
{\cal I}^j_{(a)} = C_4 \hat{D}^j U_{m{(a)}}  U^n_{(b)}  \hat{D}_n U^{m(b)}  \,.
\label{CU31}
\end{equation}
Finally, the Lagrange multiplier can be found as
\begin{equation}
\lambda =  U_j^{(a)} \left[\D_i {\cal J}^{ij}_{(a)}
- {\cal I}^j_{(a)} \right]  \,,
\label{CU4}
\end{equation}
i.e., the set (\ref{CU1}) contains $4N^2{-}5$ independent equations for the vector fields $U^k_{(a)}$ (instead of 3 ones in the U(1) model).

\subsubsection{Master equations for the gravitational field}

The gravitational field is described by the set of equations
\begin{equation}
 R_{pq} {-} \frac{1}{2} R  g_{pq} =  \Lambda g_{pq} + \lambda U^{(a)}_p  U_{(a)q}  +
 T_{pq} + T^{(\rm YM)}_{pq} \,.
\label{CE1}
\end{equation}
The stress-energy tensor of the color vector fields $T_{pq}$ is of the following form:
$$
T_{pq} = \frac12 g_{pq} {\cal K}^{ijmn}_{(a)(b)}  \hat{D}_i U^{(a)}_m \hat{D}_j U^{(b)}_n+
$$
$$
+ C_1\left[\hat{D}_m U_p^{(a)} \hat{D}^m U_{q(a)} - \hat{D}_p U^{m(a)} \hat{D}_q U_{m(a)} \right] +
$$
$$
+ C_4 U^m_{(a)} \hat{D}_m U^{(a)}_p
U^n_{(b)} \hat{D}_n U^{(b)}_q +
$$
\begin{equation}
{+} G_{(a)(b)}\D^m \left[U^{(b)}_{(p}{\cal J}_{q)m}^{(a)} {-}
{\cal J}_{m(p}^{(a)}U^{(b)}_{q)} {-}
{\cal J}_{(pq)}^{(a)} U^{(b)}_m \right].
\label{CE2}
\end{equation}
Here the parentheses $(pq)$ denote the symmetrization with respect to the coordinate indices $p$ and $q$. The symbol $T^{(\rm YM)}_{pq}$ relates to the stress-energy tensor of the Yang-Mills field
\begin{equation}
T^{(\rm YM)}_{pq} = \frac14 g_{pq}  F^{(a)}_{mn} F_{(a)}^{mn} -  F^{(a)}_{pm} F_{(a) qn} g^{mn}\,.
\label{CE4}
\end{equation}
Thus, the total set of master equations is composed by (\ref{CE1}), (\ref{CU1}), (\ref{Col1}), and (\ref{Aeq2}).

\subsection{The tensor of color polarization}

Based on the multiplet of the color fields $U_k^{(a)}$ we introduce an auxiliary quantity
\begin{equation}\label{H}
H^{(a)(b)} \equiv g^{pq} U_p^{(a)} U_q^{(b)}\,.
\end{equation}
It presents a set of scalar functions from the point of view of  spacetime transformations, and is a symmetric real tensor in the group space. The quantity $H^{(a)(b)}$ can be indicated as the tensor of color polarization, the $N^2-1$ dimensional analog of the two-dimensional polarization tensor in optics (see, e.g., \cite{LL2}). The components of this tensor depend on time and spatial coordinates. If we work with the matrix $H^{(a)}_{(b)} \equiv G_{(c)(b)} H^{(a)(c)}$ as with the algebraic quantity calculated in the fixed point of the spacetime, we can find  $N^2{-}1$ eigenvalues,
$\sigma_{\{\alpha\}}$, and the corresponding $N^2{-}1$ eigenvectors $q^{(a)}_{\{\alpha\}}$ (the subscript $\{\alpha\}$ indicates the serial number of the eigenvalue):
\begin{equation}\label{H93}
H^{(a)}_{(b)}q^{(b)}_{\{\alpha\}}  = \sigma_{\{\alpha\}} q^{(a)}_{\{\alpha\}} \,.
\end{equation}
Since the trace $H^{(a)}_{(a)}$ is equal to one due to the relationship
\begin{equation}\label{H90}
H^{(a)}_{(a)} = G_{(a)(b)}g^{mn} U^{(a)}_m U^{(b)}_n = 1  \,,
\end{equation}
the sum of eigenvalues is equal to one,
\begin{equation}\label{H905}
\sum_{\alpha =1}^{N^2-1} \sigma_{\{\alpha\}} = 1 \,.
\end{equation}
One can decompose the tensor $H^{(a)(b)}$ into the series of products of eigenvectors:
\begin{equation}\label{H97}
H^{(a)(b)} = \sum_{\alpha =1}^{N^2-1} \sigma_{\{\alpha\}}q^{(a)}_{\{\alpha\}}q^{(b)}_{\{\alpha\}}  \,.
\end{equation}
There are two physically motivated limiting cases associated with the tensor of color polarization.

1. When all the eigenvalues are equal to one another, and thus  $\sigma_{\{\alpha\}} = \sigma = 1/(N^2-1)$, we deal with the color analog of the so-called natural light \cite{LL2}.

2. When all the eigenvalues, except one, are equal to zero, we deal with the color analog of the linearly polarized light; in this case only one eigenvalue is non-vanishing, say, $\sigma_{\{1\}}$, and it is equal to one. We obtain now $H^{(a)(b)}= q^{(a)}_{\{1\}}q^{(b)}_{\{1\}}$, where $q^{(a)}_{\{1\}}$ is the corresponding eigenvector; the determinant of the matrix of this color tensor is equal to zero, the matrix is degenerated, and thus, the equality ${\rm det} H^{(a)(b)}{=}0 $ can be considered as a criterion for recognition of the {\it color polarization}.

The crucial point of the presented model is the spontaneous process,  which transforms the multiplet of color vector fields  $U^i_{(a)}$ into the bundle of vectors $q_{(a)}U^i$, which are parallel in the group space. We can monitor this process using the behavior of the eigenvalues $\sigma_{\{\alpha\}}$ of the matrix of color polarization $H^{(a)(b)}$. Our ansatz is that $N^2-2$ eigenvalues have the stepwise form
\begin{equation}\label{eigen}
\sigma_{\{\alpha\}} = \eta\left[\Omega^{*}_{\{\alpha\}}-\Omega \right] \ \sigma^{0}_{\{\alpha\}} \left[\Omega^{*}_{\{\alpha\}}-\Omega \right]^{\mu} \,.
\end{equation}
Here $\eta[z]$ is the Heaviside function equal to zero, when the argument $z$ is negative. The quantities  $\Omega^{*}_{\{\alpha\}}$, where ${\{\alpha\}} =2,3...N^2-1$, are the critical values of the  scalar $\Omega$ introduced in (\ref{A77}), (\ref{A88}). These critical values can coincide or be different. The quantity $\sigma^{0}_{\{\alpha\}}$ is a constant, and the parameter $\mu$ plays the role of critical index; we assume that $\mu>1$ providing the
function $\sigma_{\{\alpha\}}$ and its derivative to be equal to zero at the critical moment.
This stepwise representation means that there are critical moments of time $t^{*}_{\{\alpha\}}$, or equivalently, critical values of the scalar $\Omega$, for which $N^2{-}2$ eigenvalues vanish and only one of them happens to be equal to one, since the total sum has to be unit. It may be a cascade of the eigenvalue disappearances (with many different critical times), or synchronized process (with only one critical time $t_*$), however, the final of this process can be characterized by the degenerated matrix $H^{(a)(b)}$ with the rank equal to one. Thus, when $\Omega>{\rm max}\left\{\Omega^{*}_{\{\alpha\}}\right\}$, the color vector $q^{(a)}$ appears, which forms the polarization tensor $H^{(a)(b)} = q^{(a)}q^{(b)}$ according to the decomposition (\ref{H97}). Evolution of the color vector $q^{(a)}$ is described in the next Section.

\section{Phenomenology of the spontaneous color polarization}

\subsection{Basic ansatz and scheme of analysis}

\subsubsection{On the Universe evolution scenario}

We consider the toy model of the Universe expansion with three epochs, which we indicate as symmetric, interim transition, and dissymmetric, respectively, keeping in mind the analogy with the theory of ferroelectricity.

a) The evolution of the Universe during the symmetric epoch is considered to be predetermined by the coupling of the multiplet of vector fields $U^i_{(a)}$, non-Abelian Yang-Mills field with the potentials $A_k^{(a)}$, and the gravitational field.

b) The scenario of the transition stage includes three processes: first, the Yang-Mills field becomes Abelian, $A_i^{(a)} \to Q^{(a)}A_i$, decouples from the interaction with the color vector fields and degrades $A_i \to 0$; second, the procedure of color polarization reduces the color multiplet $U^i_{(a)}$ to one time-like unit vector $U^i_{(a)} \to q_{(a)}U^i$; third, the color vector $q^{(a)}$ precesses and lines up along the constant color vector $Q^{(a)}$.

c) The dissymmetric epoch is assumed to be characterized by unit vector field $U^i$, associated with the canonic aether velocity four-vector.

We focus below on the description of the transition stage only.
The anisotropic spatially homogeneous Bianchi-I model is considered to be the spacetime platform for three indicated cosmological epochs.
We monitor the state of the Universe using the color polarization tensor $H^{(a)(b)}(t)$ as a function of cosmological time. In the symmetric epoch the rank of the matrix $H^{(a)(b)}$  is assumed to be equal to $N^2{-}1$, and the determinant ${\cal H}={\rm det}H^{(a)(b)}$ to be non-vanishing. In the dissymmetric epoch the corresponding rank is equal to one, the determinant is equal to zero. We assume that during the transition epoch $N^2{-}2$ eigenvalues of the color polarization tensor (from the total $N^2{-}1$ ones) take zero values.

\subsubsection{On the parallel fields in the group space}

We assume that the mechanism of the color polarization can be adequately described in terms of fields "parallel" in the group space. This term appeared in the Yang-Mills theory \cite{Y,G,param}, and describes the quasi-Abelian gauge field configuration, for which $A^{(a)}_i {=} Q^{(a)} A_i$ with constants $Q^{(a)}$ satisfying the condition $Q^{(a)}Q_{(a)} {=}1$. For such potentials all the nonlinear terms in (\ref{46Fmn}) and in the Yang-Mills equations disappear, thus simplifying the model equations. Now we apply this ansatz to the vector field assuming that it converts into the product
\begin{equation}
U^i_{(a)}= q_{(a)}U^i   \,,
\label{ans11}
\end{equation}
but generally the color vector $q^{(a)}$ is not constant and does not coincide with the quantity $Q_{(a)}$.
The gauge covariant derivative can be now transformed into
$$
\D_m U^{(a)}_n = q^{(a)}\nabla_m U_n + U_n \nabla_m q^{(a)} +
$$
\begin{equation}
+ {\cal G} f^{(a)}_{\ (b)(c)} Q^{(b)} q^{(c)} A_m U_n \,.
\label{ans1}
\end{equation}
Clearly, when $q^{(a)}{=}Q^{(a)}$, the potential four-vector $A_i$ drops out from the derivative $\D_m U^{(a)}_n$, since the structure constant are antisymmetric.
Keeping in mind the relationship (\ref{4represent}) and assuming that the four-vector $U^i$ is unit and time-like
\begin{equation}
g_{ik} U^i U^k =1 \,,
\label{ans111}
\end{equation}
we obtain that the vector $q^{(a)}$ in the group space is also unit
\begin{equation}
G_{(a)(b)}q^{(a)} q^{(b)} = 1 \,.
\label{ans1121}
\end{equation}
Clearly, the gauge-covariant derivative $\D_m q^{(a)}$ is orthogonal to the color vector $q_{(a)}$, i.e.,
$$
0 = \frac12 \D_m [q^{(a)}q_{(a)}] = q_{(a)}\D_m q^{(a)} =
$$
$$
{=} q_{(a)}\left[\nabla_m q^{(a)} {+} {\cal G} f^{(a)}_{\ (b)(c)} A_m Q^{(b)} q^{(c)}\right] {=}
$$
\begin{equation}
= q_{(a)}\nabla_m q^{(a)} =0.
\label{1ans1}
\end{equation}
The tensor of color polarization $H^{(a)(b)}$ takes now the form
\begin{equation}
H^{(a)(b)}= q^{(a)} q^{(b)} \,.
\label{ans109}
\end{equation}
Clearly, the rank of the matrix $H^{(a)}_{(b)}$ is equal to one, and the determinant is equal to zero.
The color vector $q^{(a)}$ is an eigenvector of $H^{(a)}_{(b)}$ with the unit eigenvalue:
\begin{equation}
H^{(a)}_{(b)} q^{(b)} = q^{(a)}\,.
\label{ans109b}
\end{equation}
In other words, the mechanism of parallelization in the group space applied to the color vectors $U^i_{(a)}$ gives the result analogous to the polarization of light in optics.
Indeed, one can say that a special direction in the group space appears, which is associated with the color vector $q^{(a)}$, along which all the color vectors $U^i_{(a)}$ are lined up.
Generally, $q^{(a)}(t)$ precesses in the group space, but when the color vector $q^{(a)}$ becomes constant, one can transform the matrix $H^{(a)(b)}$ into ${\rm diag}\{1,0,0...0 \}$ for arbitrary time moment, using the admissible rotation in the group space. Description of the evolution of the color vector $q^{(a)}(t)$ is the crucial point of our analysis.

\subsubsection{On the spacetime platform}

For the illustration of the suggested idea, we consider the class of homogeneous spacetimes with the metric
\begin{equation}
ds^2 = dt^2 - a^2(t)dx^2 - b^2(t)dy^2 - c^2(t)dz^2  \,.
\label{Bianchi}
\end{equation}
Generally, this metric describes the Bianchi-I model. When $a(t){=}b(t){=}c(t)$ we obtain the Friedmann type model. We assume that all the unknown model state functions inherit the spacetime symmetry and depend on the cosmological time only.

The global unit timelike vector $U^i$ is assumed to be of the form $U^i=\delta^i_t$; below we show that the model admits this construction. The metric (\ref{Bianchi}) provides the covariant derivative to be symmetric, the acceleration  four-vector $a_i$ to vanish, and the scalar of expansion $\Theta$ to have very simple form
$$
\nabla_m U_n = \nabla_n U_m = \frac12 \dot{g}_{mn}   \,, \quad  a_i \equiv U^k \nabla_k U_i =0 \,,
$$
\begin{equation}
\Theta(t) \equiv \nabla_k U^k = \frac{\dot{a}}{a} + \frac{\dot{b}}{b} +\frac{\dot{c}}{c} \,.
\label{B2}
\end{equation}
Here and below the dot denotes the ordinary derivative with respect to time. In addition, we use two differential consequences of the normalization conditions: $U^k \nabla_i U_k =0$, and $q_{(c)}U^k \nabla_k q^{(c)} =0$.
The listed assumptions simplify the model essentially, and we start to discuss its details.

\subsection{Reduced master equations}

\subsubsection{The Yang-Mills field in the transition epoch}

As we already noticed, we assume that the first manifestation of the phase transition in the model under discussion is connected with the parallelization of the Yang-Mills potentials, i.e., $A_i^{(a)} \to Q^{(a)} A_i$, where $Q^{(a)}$ is a constant vector in the group space; the second manifestation is the parallelization of the color vector fields $U_i^{(a)} \to q^{(a)} U_i$. For such fields, parallel in the group space, the strength of the Yang-Mills field (\ref{46Fmn}) has the quasi-Maxwellian form
\begin{equation}
F^{(a)}_{mn} = Q^{(a)}[\nabla_m A_n - \nabla_n A_m ]\,. \label{046Fmn}
\end{equation}
Based on the Bianchi-I spacetime platform and taking into account the antisymmetric properties of the structure constants, we find the color current $\Gamma^i_{(a)}$ produced by the color vector fields:
$$
\Gamma^i_{(a)} = \frac{{\cal G}}{\kappa}  (C_1{+}C_2{+}C_3) U^i f_{(d)(c)(a)} q^{(c)} \times
$$
$$
\times \left[\dot{q}^{(d)} - {\cal G}f_{(d)(b)(h)}Q^{(b)}q^{(h)} U^m A_m \right]  -
$$
\begin{equation}\label{gamma0}
{-} \frac{{\cal G}^2}{\kappa} C_1 A^k \Delta_k^i f^{(d)}_{(c)(a)} q^{(c)}f_{(d)(b)(h)}Q^{(b)}q^{(h)}.
\end{equation}
Here $\Delta^i_k \equiv \left(\delta^i_k - U^i U_k \right)$ is the projector. Clearly, as long as, the color current is not vanishing, $\Gamma^i_{(a)} \neq 0$, or is not linear in the potential $A_i$, the Yang-Mills equations (\ref{Col1}) do not admit the trivial solutions $A_i=0$. In principle, this color current disappears, when $q_{(a)}=Q_{(a)}=const$; we consider this state as the final one. When $q^{(a)}$ depends on time and thus $\dot{q}^{(a)}\neq 0$, one can obtain the trivial solution $A_i=0$, if to use the special condition for the Jacobson's coupling constants
\begin{equation}\label{gamma1}
C_1 + C_2 + C_3 = 0 \,,
\end{equation}
which was motivated, e.g., in  \cite{CCC}. For this case the color current is proportional to the potential $A^k$, and the trivial exact solution $A^i=0$ exists. This means that the system can choose this trivial solution for the Yang-Mills field in the bifurcation associated with the end of the first phase transition. Of course, the condition (\ref{gamma1}) narrows down the frameworks of the canonic Einstein - aether theory, however, this restricted model is very illustrative for our purpose.

\subsubsection{Solutions to the equations for the color vector fields}

Now we can consider the reduced master equations for the multiplet of vector fields with the conditions $A^i_{(a)}{=}0$ and $C_1{+}C_2{+}C_3{=}0$.
Clearly,  the vectorial quantity ${\cal I}^j_{(a)}$   (\ref{CU31}) vanishes, since the following vector is equal to zero:
\begin{equation}
U^n_{(b)} \D_n U^{m (b)} {=} q_{(b)}q^{(b)}U^n \nabla_n U^m {+} U^m q_{(b)} \dot{q}^{(b)}.
\label{00CU2}
\end{equation}
The tensor  ${\cal J}^{ij}_{(a)}$ (\ref{CU2}) happens to be proportional to the coupling constant $C_2$:
\begin{equation}
{\cal J}^{ij}_{(a)} {=} C_2 \left\{q_{(a)}\left[g^{ij} \Theta {-} \nabla^{(i} U^{j)}  \right] {+} \dot{q}_{(a)} \Delta^{ij}\right\} .
\label{0CU2}
\end{equation}
Then we see that $3(N^2{-}1)$ equations from the $4(N^2{-}1)$ ones, written in (\ref{CU1}) convert into the trivial identities $0=0$; only $N^2{-}1$ equations for $j=0$ have the nontrivial form
\begin{equation}
q_{(a)} C_2 \left(\frac{\ddot{a}}{a} + \frac{\ddot{b}}{b} + \frac{\ddot{c}}{c} \right) = \lambda q_{(a)}  \,.
\label{0CU1}
\end{equation}
All these equations are satisfied simultaneously, if the Lagrange parameter
$\lambda(t)$ is chosen to be of the form:
\begin{equation}
 \lambda  =  C_2 \left(\frac{\ddot{a}}{a} + \frac{\ddot{b}}{b} + \frac{\ddot{c}}{c} \right)   \,.
\label{0CU11}
\end{equation}
We have to emphasize that in the presented model the master equations for the multiplet of vector fields do not impose any restrictions on the color vector $q^{(a)}(t)$. This means that we can suggest our version of the model of the color vector $q^{(a)}(t)$ evolution.

\subsubsection{Gravity field equations}

To explain the details of the structure of the reduced aether stress-energy tensor (\ref{CE2})
we have to make four preliminary remarks.

1. Using (\ref{0CU2}) and the relationships
\begin{equation}
\nabla_i(q^{(a)} U_j) = \frac12 q^{(a)}\dot{g}_{ij} {+} \dot{q}^{a} U_iU_j
\label{aux1}
\end{equation}
we can calculate the scalar
$$
{\cal K}^{ijmn}_{(a)(b)}  \hat{D}_i U^{(a)}_m \hat{D}_j U^{(b)}_n  {=} C_2 \left[\Theta^2 {-} \nabla_i U_j \nabla^i U^j \right] {=}
$$
\begin{equation}
= 2 C_2 \left[\frac{\dot{a}}{a} \frac{\dot{b}}{b} {+} \frac{\dot{a}}{a} \frac{\dot{c}}{c} {+} \frac{\dot{b}}{b} \frac{\dot{c}}{c} \right] \,.
\label{j}
\end{equation}

2. Since the tensor $\nabla_i U_j$ is symmetric, the term proportional to $C_1$ in (\ref{CE2}) vanishes.

3. Since $U^k \nabla_k U_j =0$ and $q_{(a)} \dot{q}^{(a)}=0$, the term with $C_4$ in (\ref{CE2}) also vanishes.

4. Since ${\cal J}^{(a)}_{ij}={\cal J}^{(a)}_{ij}$, the last term in (\ref{CE2})
converts into
\begin{equation}
C_2\left[{-}g_{pq} \left(\dot{\Theta}+\Theta^2 \right) {+} \frac12 \Theta \dot{g}_{pq} {+} U^m \nabla_m (\nabla_p U_q)  \right].
\label{jj}
\end{equation}
Thus, for the tensor $T^p_q$ we obtain the formula, which surprisingly contains neither the color vector $q^{(a)}$, nor its derivative $\dot{q}^{(a)}$:
$$
T^p_q {=} C_2 \left\{{-}\delta^p_q  \left(
\frac{\ddot{a}}{a} {+}  \frac{\ddot{b}}{b} {+}  \frac{\ddot{c}}{c} {+}
\frac{\dot{a}}{a} \frac{\dot{b}}{b} {+} \frac{\dot{a}}{a} \frac{\dot{c}}{c} {+} \frac{\dot{b}}{b} \frac{\dot{c}}{c} \right)  {+} \right.
$$
$$
\left. {+} \delta^p_1 \delta^1_q \left(\frac{\ddot{a}}{a} {+} \frac{\dot{a}}{a} \frac{\dot{b}}{b} {+} \frac{\dot{a}}{a} \frac{\dot{c}}{c} \right)
 {+} \delta^p_2 \delta^2_q \left(\frac{\ddot{b}}{b} {+} \frac{\dot{b}}{b} \frac{\dot{c}}{c} {+} \frac{\dot{a}}{a} \frac{\dot{b}}{b} \right) {+} \right.
$$
\begin{equation}
\left. +  \delta^p_3 \delta^3_q \left(\frac{\ddot{c}}{c} + \frac{\dot{a}}{a} \frac{\dot{c}}{c} + \frac{\dot{b}}{b} \frac{\dot{c}}{c} \right) \right\} \,.
\label{jjj}
\end{equation}
Taking into account the term for $\lambda$ (\ref{0CU11}), we can now represent four non-trivial gravity field equations in the following form:
$$
\left[\frac{\dot{a}}{a} \frac{\dot{b}}{b} + \frac{\dot{a}}{a} \frac{\dot{c}}{c} + \frac{\dot{b}}{b} \frac{\dot{c}}{c} \right] (1+C_2)= \Lambda  \,,
$$
$$
\left[\frac{\ddot{b}}{b} +  \frac{\ddot{c}}{c} + \frac{\dot{b}}{b} \frac{\dot{c}}{c} \right] (1+C_2) = \Lambda  \,,
$$
$$
\left[\frac{\ddot{a}}{a} +  \frac{\ddot{c}}{c} + \frac{\dot{a}}{a} \frac{\dot{c}}{c}\right] (1+C_2)  = \Lambda \,,
$$
\begin{equation}\label{Ein}
\left[\frac{\ddot{b}}{b} +  \frac{\ddot{a}}{a} + \frac{\dot{b}}{b} \frac{\dot{a}}{a}\right] (1+C_2)  = \Lambda  \,.
\end{equation}
It is the well known system of equations, the novelty is only in the structure of the constant: we deal now with the modified cosmological constant $\frac{\Lambda}{1+C_2}$ instead of $\Lambda$.
The  symmetry of the equations (\ref{Ein}) allows us to find explicitly the expansion scalar $\Theta(t)$ as the function of time. Indeed, if we calculate the auxiliary function $\dot{\Theta} {+} \Theta^2$, we obtain
\begin{equation}\label{thetaj}
\dot{\Theta} + \Theta^2 = \frac{\ddot{a}}{a} + \frac{\ddot{b}}{b} + \frac{\ddot{c}}{c} + 2\left[\frac{\dot{a}}{a} \frac{\dot{b}}{b} + \frac{\dot{a}}{a} \frac{\dot{c}}{c} + \frac{\dot{b}}{b} \frac{\dot{c}}{c} \right]
\,.
\end{equation}
The appropriate linear combination of the equations (\ref{Ein}) shows that the right-hand side of (\ref{thetaj}) is equal to $\frac{3\Lambda}{1+C_2}$. In other words, we are faced with the equation
\begin{equation}\label{theta1j}
\dot{\Theta} + \Theta^2 = \frac{3\Lambda}{1+C_2}
\,,
\end{equation}
the solution to which is the following:
\begin{equation}\label{theta2}
\Theta(t) = 3H_0 \left\{ \frac{\Theta(t_0) {+} 3H_0 \ {\rm th}[3H_0(t{-}t_0)]}{\Theta(t_0) \ {\rm th}[3H_0(t{-}t_0)] {+} 3H_0 } \right\} \,,
\end{equation}
where the parameter $H_0$ is introduced as
\begin{equation}\label{theta25}
H_0 \equiv \sqrt{\frac{\Lambda}{3(1+C_2)}}\,.
\end{equation}
The function $\Theta(t)$ starts from the value $\Theta(t_0)$ at $t=t_0$ and finishes with the value $3H_0$ at $t \to \infty$. When $|\Theta(t_0)|<3H_0$, the function $\Theta(t)$ is regular in the interval $t_0<t<\infty$, and is monotonic with $\dot{\Theta}>0$. When $\Theta(t_0)=3H_0$, the solution is constant, $\Theta(t)=3H_0$.

Also, one can find the explicit formula for the unit volume in the three-dimensional space
\begin{equation}\label{gamma11}
V(t)= \frac{a(t) \cdot b(t) \cdot c(t)}{a(t_0)b(t_0)c(t_0)} \,,
\end{equation}
for which $V(t_0)=1$, and
\begin{equation}\label{gamma12}
\frac{\dot{V}}{V}= \frac{\dot{a}}{a} + \frac{\dot{b}}{b} + \frac{\dot{c}}{c} = \Theta(t) = \nabla_k U^k \,.
\end{equation}
Using (\ref{theta2}) and (\ref{gamma12}), we obtain now
\begin{equation}\label{theta3}
V(t) = \frac{\Theta(t_0)}{3H_0}  {\rm sh}[3H_0(t{-}t_0)] {+} {\rm ch}[3H_0(t{-}t_0)] \,.
\end{equation}
Finally, it is clear that the differences of logarithmic derivatives have the form
\begin{equation}\label{theta33}
\frac{\dot{b}}{b} {-} \frac{\dot{a}}{a} = \frac{K_1}{V(t)} , \quad
\frac{\dot{c}}{c} {-} \frac{\dot{b}}{b} = \frac{K_2}{V(t)} , \quad\frac{\dot{a}}{a} {-} \frac{\dot{c}}{c} = \frac{K_3}{V(t)} ,
\end{equation}
where $K_1$, $K_2$, $K_3$ are the integration constants, which satisfy the condition $K_1{+}K_2{+}K_3 {=} 0 $. To make sure, that (\ref{theta33}) take place, we can check, e.g., that due to the second and third equations from the set (\ref{Ein})
\begin{equation}\label{theta373}
\frac{d}{dt}\left(\frac{\dot{b}}{b} {-} \frac{\dot{a}}{a}\right) = - \left(\frac{\dot{b}}{b} {-} \frac{\dot{a}}{a} \right) \Theta \,,
\end{equation}
and then integrate this equation.
Asymptotically, at $V \to \infty$, the rates of expansion $\frac{\dot{a}}{{a}}$, $\frac{\dot{b}}{{b}}$ and $\frac{\dot{c}}{{c}}$ coincide, i.e., the Universe tends to be spatially isotropic.

\subsection{Dynamics of the color vector $q^{(a)}(t)$}

\subsubsection{Exact solution to the precession equation}

According to our scenario, after the phase transition, when the specific direction $Q^{(a)}$ in the group space appeared, and the color vector fields have become parallel, $U^{i}_{(a)}{=}q_{(a)}U^i$, the process of precession of the color vector $q^{(a)}$ around the color vector $Q^{(a)}$ was started up. This process can be described by the  equation, which is the generalization of the Wong equation for the color charges \cite{Wong}
$$
U^m \D_m q^{(a)} = {\cal F}^{(a)} \,,
$$
\begin{equation}\label{gamma5}
{\cal F}^{(a)} = \nu \left[\delta^{(a)}_{(b)}- q^{(a)}q_{(b)} \right]Q^{(b)} \,.
\end{equation}
In the Bianchi-I spacetime platform, these equations can be presented in more detailed form:
$$
\dot{q}^{(a)} + {\cal G} f^{(a)}_{\ (b)(c)} Q^{(b)} q^{(c)} (U^m A_m) =
$$
\begin{equation}\label{gamma58}
= \nu(t) \left[Q^{(a)} - q^{(a)}q_{(b)} Q^{(b)} \right] \,,
\end{equation}
where $\nu(t)$ is some function of time.
The force-like term ${\cal F}^{(a)}$ in (\ref{gamma5}) is orthogonal to the color vector $q_{(a)}$: i.e., $q_{(a)}{\cal F}^{(a)} {=}0$. This property of the force-like term ${\cal F}^{(a)}$ guaranties that the normalization condition $q_{(a)}q^{(a)}=1$ is supported. The second term in the left-hand side of the equation (\ref{gamma58}) is also orthogonal to the color vector $q_{(a)}$; it  can be eliminated using the so-called Landau's gauge condition for the Abelian potential four-vector, $U^iA_i=0$. The equations, modified correspondingly,
\begin{equation}\label{gamma581}
\dot{q}^{(a)} = \nu(t) \left[Q^{(a)}- q^{(a)}q_{(b)} Q^{(b)} \right] \,,
\end{equation}
look like the equations of motion of the macroscopic particle with unit mass under the influence of the Stokes-type force in the fluid flow, characterized by the velocity $Q^{(a)}$.
In order to solve the equation (\ref{gamma581}) we consider the scalar product $X(t) \equiv q_{(a)}(t)Q^{(a)}$; since $q^{(a)}(t)q_{(a)}(t)=1$ and $Q^{(a)}Q_{(a)}=1$, the scalar product $X$ is equal to the cosine of the angle between the vectors $q^{(a)}$ and $Q^{(b)}$ in the group space. Then we find its derivative $\dot{X}= Q_{(a)}{\cal F}^{(a)}$, and obtain the equation, which is very famous in the theory of differential equations:
\begin{equation}\label{gamma6}
\dot{X} = \nu(t) (1- X^2) \,.
\end{equation}
This equation admits two special constant solutions $X {=} {\pm} 1$, which relate to the cases $q^{(a)}(t){=} {\pm} Q^{(a)}$, i.e., $q^{(a)}$ and $Q^{(a)}$ are parallel or anti-parallel, respectively, for arbitrary moment of the cosmological time. For arbitrary initial value $X(t_0)$ the solution to the equation (\ref{gamma6}) is known to be of the following form:
$$
X(t) = \frac{e^{2 T(t)}(1+X(t_0))- (1-X(t_0))}{e^{2 T(t)}(1+X(t_0))+ (1-X(t_0))} \,,
$$
\begin{equation}\label{gamma7}
T(t) \equiv \int_{t_0}^t d\tau \nu (\tau) \,.
\end{equation}
Returning to the equation (\ref{gamma581}) rewritten in the form
\begin{equation}\label{gamma51}
\frac{\ d q^{(a)}}{d T}  = Q^{(a)} - q^{(a)}X(T) \,,
\end{equation}
we readily find the solution
$$
q^{(a)}(t)  = \frac{q^{(a)}(t_0)}{\left[{\rm ch}T + X(t_0){\rm sh}T \right]} +
$$
\begin{equation}\label{gamma52}
+ Q^{(a)} \frac{\left[{\rm sh}T + X(t_0)\left({\rm ch}T -1\right) \right]}{\left[{\rm ch}T + X(t_0){\rm sh}T \right]} \,.
\end{equation}
For illustration, it is convenient to consider the case with $X(t_0){=}0$; it corresponds to the configuration with initially orthogonal color vectors $q^{(a)}(t_0)$ and $Q^{(a)}$; now we obtain
$$
X(t) = {\rm th}T \,,
$$
\begin{equation}\label{gamma9}
q^{(a)}(t)  = \frac{q^{(a)}(t_0)}{{\rm ch}T} + Q^{(a)}{\rm th}T \,,
\end{equation}
$$
H^{(a)(b)}(t) = \frac{H^{(a)(b)}(t_0)}{{\rm ch}^2T} + Q^{(a)} Q^{(b)}{\rm th}^2T  +
$$
\begin{equation}\label{gamma97}
+ \left[Q^{(a)}q^{(b)}(t_0)+ Q^{(b)}q^{(a)}(t_0) \right] \frac{{\rm sh}T}{{\rm ch}^2T} \,.
\end{equation}
Asymptotically, at $T \to \infty$ we obtain $q^{(a)} \to Q^{(a)}$. In other words, during the evolution process the color vector $q^{(a)}$ forgets its initial value $q^{(a)}(t_0)$ and is forced to be lined up along the direction $Q^{(a)}$. The question arises: what is the rate of asymptotic parallelization of the vectors $q^{(a)}$ and $Q^{(d)}$. Clearly, this process is predetermined by the  properties of the function $\nu(t)$; let us consider an example of its modeling.

\subsubsection{Model function $\nu(t)$ }

For the modeling of the function $\nu(t)$ we use the following construction:
\begin{equation}\label{gamma71}
\nu (t) = 2 \gamma T_{*} \Theta(t) \frac{V^2}{\left[V^2_{*}-V^2(t) \right]^{1+\gamma}} \,,
\end{equation}
and calculate the function  $T(t)$ (\ref{gamma7}):
\begin{equation}\label{gamma8}
T(t) = T_{*} \left\{\frac{1}{\left[V^2_{*}{-} V^2(t)\right]^{\gamma}}{-} \frac{1}{\left[V^2_{*}{-} 1\right]^{\gamma}} \right\}\,.
\end{equation}
Here $T_{*}$ and $\gamma$ are positive parameters.
Let us make three remarks concerning this formula.

1. When the function $V(t)$ grows, it reaches the value $V_{*}$, so,  at the finite moment of the cosmological time $t=t_{*}$ the function $T(t)$ reaches infinity, $T(t_{*}){=} \infty$.

2. According to (\ref{gamma52}), at $T(t_{*})=\infty$ the color vector $q^{(a)}(t_*)$ coincides with $Q^{(a)}$, i.e., the process of their parallelization finished during the finite time interval.

3. At the moment $t=t_{*}$  the derivative of the color vector $q^{(a)}$ takes zero value, $\dot{q}^{(a)}(t_{*})=0$.

In order to illustrate the last statement, we calculate the derivative of the function $q^{(a)}(t)$ presented in (\ref{gamma9}) for the case $X(t_0)=0$:
\begin{equation}\label{gamma99}
\dot{q}^{(a)}(t)  = \frac{\nu(t)}{{\rm ch}T} \left[\frac{Q^{(a)}}{{\rm ch}T} - q^{(a)}(t_0) {\rm th}T  \right] \,.
\end{equation}
For big values of the function $T$ one can use the following replacement (compare (\ref{gamma71}) and (\ref{gamma8})):
\begin{equation}\label{gamm0a94}
\nu(t)  \to 2 \gamma T^{-\frac{1}{\gamma}}_{*} \Theta V^2(t_{*})T^{1+\frac{1}{\gamma}} \,,
\end{equation}
where $\Theta$ (\ref{theta2}) is the regular finite function. Since for arbitrary $\alpha$ the limit $\lim_{T \to \infty} \left[\frac{T^{\alpha}}{{\rm ch}T}\right]$ is equal to zero, we can confirm the condition $\dot{q}^{(a)}(t_{*}){=}0$.
The critical time moment $t_{*}$, for which $V(t_*){=}V_{*}$ can be found as
\begin{equation}\label{theta5}
3H_0(t_* - t_0) =
\end{equation}
$$
={\rm ln} \left\{\frac{3H_0 V_{*}}{3H_0{+}\Theta(t_0)} \left[1 {+} \sqrt{1{+} \frac{\Theta^2(t_0){-}9H_0^2}{9H_0^2 V^2_{*}}}\right] \right\}.
$$
One can check that $t_{*}> t_0$, when $V_{*}>1$. For the special case, when $3H_0{=}\Theta(t_0){=}\Theta(t)$, we obtain $t_* = t_0 {+} \frac{1}{3H_0} {\rm ln} V_{*}$.

\subsubsection{Model threshold function $\Omega$}

Let us consider now the function $\Omega$ introduced by (\ref{A77}) and (\ref{A88}), the value of which indicates the nearness of the event of the spontaneous color polarization.
We have now that
\begin{equation}\label{Omega}
\Omega^{(a)} = \dot{q}^{(a)} + q^{(a)} \Theta  \,, \quad \Omega^2 =  \dot{q}^{(a)}\dot{q}_{(a)} + \Theta^2 \,,
\end{equation}
thus, for the illustrative function (\ref{gamma9}) we obtain
\begin{equation}\label{Omega1}
\Omega =  \sqrt{\frac{\nu^2}{{\rm ch}^2T} + \Theta^2} \,.
\end{equation}
Taking into account (\ref{gamma71}) and (\ref{gamma8}), we finally reconstruct the function $\Omega$ as
\begin{equation}\label{Omega2}
\Omega = \Theta \ \sqrt{1{+} \frac{4\gamma^2 T^2_{*} V^4}{{\rm ch}^2T} \left[\frac{T}{T_{*}}{+} \frac{1}{\left(V^2_{*}{-}1\right)^{\gamma}} \right]^{\frac{2(1{+}\gamma)}{\gamma}} }\,.
\end{equation}
Near the critical point $t=t_{*}$, when $T \to \infty$, the threshold function $\Omega$ behaves as the expansion scalar $\Theta$, i.e., the summarized rate of the Universe expansion
$\Theta = \frac{\dot{a}}{a}+ \frac{\dot{b}}{b}+\frac{\dot{c}}{c}$ predetermines the rate of the onset of the spontaneous color polarization.

\section{Discussion and conclusion}

We share the point of view of many cosmologists, that a plethora of fields and particles in the early hot Universe were lost, destroyed or transformed as a result of a series of phase transitions; now we observe  a limited number of fundamental fields and stable elementary particles. Our goal is to reconstruct the history of emergence of the cosmic substratum indicated as dynamic aether; to be more precise, we are interested to establish a model of formation of the unit timelike vector field as a splinter of the family of vector fields from the SU(N) symmetric multiplet. Our phenomenological model is based on three elements.

1) The first element of the model is the procedure of the spontaneous polarization of the SU(N) multiplet of vector fields. This procedure is described in terms of critical behavior of the eigenvalues of the color polarization tensor (see (\ref{H}), (\ref{H97}) and (\ref{eigen})). When the Universe expands, and the scalar $\Omega$ reaches the critical value $\Omega^*_{\{\alpha \}}$, the corresponding eigenvalue with the serial number ${\alpha}$ vanishes. When $\Omega>{\rm max}\left\{\Omega^{*}_{\{\alpha\}}\right\}$, we obtain the situation with only one non-vanishing eigenvalue; this last eigenvalue is equal to one, since the trace of the polarization tensor is unit. Based on the analogy with optics, we have to state now, that we deal with linearly polarized system, i.e., the color polarization took place. Mathematically, this procedure means that all the vectors from the SU(N) symmetric multiplet have become parallel in the group space to one vector $U^i$, namely, $U^i_{(a)}= q_{(a)}U^i$.

2) The second element of the model is the mechanism of precession of the color vector $q^{(a)}$ around the constant color vector $Q^{(a)}$ formed in course of parallelization of the Yang-Mills potential $A^{(a)}_i \to Q^{(a)} A_i$. Evolution of the color vector $q^{(a)}$ was described by the equation of the Wong type (\ref{gamma5}); the exact solution to this equation is presented in (\ref{gamma52}). Asymptotically, $q^{(a)}(t)$ tends to the constant $Q^{(a)}$, and the system as a whole becomes quasi-Abelian, since all the structure constants $f_{(a)(b)(c)}$ fall out from the basic formulas.

3) The third element of the model is the calculation of the rate of the precession of the color vector $q^{(a)}$. We have chosen the guiding function $\nu(t)$ in (\ref{gamma58}) in the critical form (\ref{gamma71}) providing the process of parallelization of $q^{(a)}$ and  $Q^{(a)}$ to be finished during the finite time interval (see (\ref{gamma8}) and (\ref{gamma52})).

The conclusion of our study is the following: the aether velocity four-vector can be considered as a rudiment, which remained after the spontaneous color polarization accompanying the phase transition in the early Universe.

\vspace{5mm}
\noindent
{\bf Acknowledgments}

\noindent
The work was supported by the Russian Foundation for Basic Research (Project No. 20-02-00280), and by the Program of Competitive Growth of Kazan Federal University.

\end{document}